\documentstyle[preprint,eqsecnum,aps,prb]{revtex}
\tightenlines
\begin{document}
\draft
\title{Magnetoplasma excitations of two vertically coupled dots}
\author{B. Partoens\cite{bart}, A. Matulis\cite{algis}, and F. M. Peeters\cite 
{francois}}
\address{Departement Natuurkunde, Universiteit Antwerpen (UIA),\\
Universiteitsplein 1, B-2610 Antwerpen, Belgium}
\date{\today}
\maketitle

\begin{abstract}
A classical hydrodynamic approach is used to calculate the magnetoplasma
excitations of two vertically coupled electron dots. The electrons are
confined by different parabolic potentials in which case Kohn's theorem is no
longer valid. The
equilibrium density profiles of the electrons in both dots are calculated 
as function of
the interdot distance. 
We find that for unequal confinements of the two dots the electron density in
one of the dots becomes ring-like.
The electron densities are then used to obtain the
magnetoplasma frequencies. The oscillator
strengths are calculated, and we find that other then the center of
mass modes can be excited due to the electron-electron interactions.

\end{abstract}


\section{Introduction}
\newcommand{\tens}[1]{{\stackrel{\leftrightarrow}{#1}}}
Modern semiconductor technology makes it possible to fabricate low-dimensional
nanostructures with controllable chemical composition and geometric structure.
In artificial atoms a number of electrons are confined into a single
quasi-two-dimensional quantum dot which has been widely studied over the past
few years
(see, for instance, Ref.~\onlinecite{johnson95}). Most of the work has focused
on quantum
dots with a soft lateral confinement potential in the heterostructure
plane
(2D-plane) which are the so called parabolic quantum dots. Such quantum
dots exhibit a far-infrared
absorption spectrum which reflects only the center of mass motion of the
electron system which is a consequence of a generalized Kohn's
theorem.~\cite{kohn} 
When a magnetic field is applied 
perpendicular to the plane of the quantum dot
it results into two absorption peaks whose positions are independent of the
number of electrons in the dot. Note that the same absorption spectrum 
follows from a pure classical analysis.
In order to reveal effects due to electron-electron interaction 
the confinement potential should be anharmonic,
or band non-parabolicity effects should be taken into account. 

The other possibility to trace the influence
of electron-electron interaction on the quantum dot properties is to
consider the more sophisticated system of coupled quantum dots. An example of
such a system are two vertically coupled dots.
The properties of such a system for small number of electrons were
studied previously via exact diagonalization~\cite{palacios95} or for a 
model system with an exactly
solvable interaction~\cite{benjamin95} where the main attention was on
electron correlations in the ground state, magic numbers and the
possibility of the appearance of new type of entangled electronic states.
In Ref.~\onlinecite{maksym} it was shown that in the case of a double dot
system, containing three electrons, with
different parabolic
confinement potentials, the electron correlation effects exhibit themselves in the
absorption energy. 
The classical double dot system containing a finite number of electrons exhibits
a rich variety of first, second and higher order transitions.~\cite{partoens97}
Two vertically aligned dots is the finite system analog of the well studied
bilayer systems~\cite{bilayers} which have been shown to exhibit much
richer physical phenomena then 2D and 3D systems. 
\par
Such coupled dot systems
can be realized for example in vertically etched double quantum well
structures,~\cite{austin} parallel layers of self assembled quantum
dots~\cite{self} and by using a conducting atomic force microscope tip 
above a quantum well
which can induce locally a dot structure in the quantum well and one
beneath the lowest barrier.~\cite{allen}
Recently, the successful growth of vertically aligned dots of up to 10
layers
of InAs islands separated by GaAs spacer layers were reported.~\cite{solomon} 
These experimental attempts are
still in their infancy and to our knowledge no magneto-absorption
experiments have been performed yet.

The purpose of the present paper is to study the properties of the system
consisting
of two vertically coupled quantum dots in a uniform perpendicular 
magnetic field. We calculate the electron density profiles,
the spectrum of magnetoplasma excitations and the corresponding
power absorption.
We restrict our consideration to dots with a large number of electrons.
In
this case the classical hydrodynamic approach is adequate, and we follow
the ideas presented in the paper of Ye and Zaremba~\cite{ye94} where 
the single quantum dot with quartic confining potential corrections was
considered.
A similar technique was used to study one layer electron dots with
parabolic\cite{nazin89,shikin91} and non-circular\cite{tevosyan92} confinement
potentials.

\section{Model}
\label{s2}

We consider two infinitely thin two-dimensional circular symmetric dots
which are vertically aligned
and are separated by a distance $a$ as is schematically shown in
Fig.~\ref{layout}. The electrons 
are confined by different parabolic potentials.
In the hydrodynamic approach the electrons are described by density
functions $\rho_j({\bf r})$ where $j=1,2$ refers to the corresponding dot,
and ${\bf r}=(x,y)$ is the position coordinate
in the plane of the dot. 
Those electron densities create an electrostatic potential 
$\varphi({\bf r},z)$ which obeys Poisson's equation
\begin{equation}\label{poisson}
  \nabla^2\varphi = \frac{4\pi e}{\epsilon}\{\rho_1({\bf r})\delta(z)
               + \rho_2({\bf r})\delta(z-a)\},
\end{equation}
where $\epsilon$ is the dielectric constant of the medium the electrons are
moving in. We assume that $\epsilon$ is constant thrroughout the whole system
which is a good approximation for e.g. the $\mbox{GaAs/Al}_{x}
\mbox{Ga}_{1-x}\mbox{As}$ system.
\par
In a purely classical approach,
the electrons in both dots achieve an equilibrium
distribution in which the lateral force arising from the confining potentials is
balanced by the electrostatic repulsive force due to the electrons.
Equivalently, the equilibrium conditions can be expressed as the constancy of
the chemical potential in each of the dots given by the expressions
\begin{mathletters}
\label{cond}
\begin{eqnarray}
  \mu_1&=&\frac{k_1}{2}r^2-e\varphi({\bf r},z)|_{z=0},\\
  \mu_2&=&
  \frac{k_2}{2}r^2-e\varphi({\bf r},z)|_{z=a}.
\end{eqnarray}
\end{mathletters}
Here the strength of the parabolic confinement potential
is characterized by the corresponding factor $k_i$ which is related to the
confinement frequency as $\omega_j=\sqrt{k_j/m}$. Here and further the symbol $m$
stands for the effective electron mass. The difference in the above
chemical potentials can be controlled by applying some external bias potential
\begin{equation}\label{bias}
  eV=\mu_1-\mu_2.  
\end{equation}
\par
In equilibrium the electron densities are circular symmetric 
$\rho_j({\bf r})=\rho_j(r)$ and vanish outside some
radius $R_j$ which must be calculated self-consistently. Those radii are related
to the number of electrons in the dots
\begin{equation}\label{number}
  2\pi\int_0^{R_j}\rho_j(r)rdr=N_j.
\end{equation}
Condition (\ref{bias}) together with definitions (\ref{cond}) and  
Poisson's equation (\ref{poisson}) form the basic set of equations  
from which the equilibrium electron density profiles in the dots can be obtained.
In a classical description the equilibrium electron distributions are not influenced by an
external magnetic field.
\par
In order to obtain the excitation spectrum and the power absorption in the
coupled dot system the small harmonic deviations from the
electron equilibrium density and potential functions have to be introduced
\begin{equation}
  \delta\rho_j({\bf r}),\;\delta\varphi({\bf r}) \sim e^{i\omega t}.
\end{equation}
Those additional quantities obey a set of dynamic equations.
They are composed of the same Poisson
equation (\ref{poisson}) for the above deviations and the continuity equations
\begin{equation}\label{cont}
  i\omega e\delta\rho_j({\bf r}) + \nabla \cdot {\bf j}_j({\bf r}) = 0.
\end{equation}
Usually the currents in the dots ${\bf j}_j({\bf r})$ are defined by solving the
linearized hydrodynamic equations for the averaged electron velocities. The solution
of these equations leads to Ohm's law
\begin{equation}\label{ohm}
  {\bf j}_j({\bf r})=-\tens{\sigma}^{(j)} \cdot \nabla \delta\varphi({\bf r},z)|_{z=0,a}.
\end{equation}
In the presence of an external applied perpendicular magnetic field, the components 
of the conductivity  tensor $\tens{\sigma}^{(j)}$ in  polar coordinates are given by
\begin{eqnarray}\label{conduct}
\sigma_{rr}^{(j)}=\sigma_{\theta\theta}^{(j)} &=& \frac{i\omega e^2 \rho_j(r)}
{m(\omega^2-\omega_c^2)}, \nonumber \\
\sigma_{r\theta}^{(j)} =\sigma_{\theta r}^{(j)} &=& \frac{\omega_c e^2 \rho_j(r)}
{m(\omega^2-\omega_c^2)}.
\end{eqnarray}
Here the symbol $\omega_c=eB/mc$ stands for the electron cyclotron frequency,
and $\omega$ is the frequency of the external perturbation,
i.e. the far-infrared radiation.
Poisson's equation together with Eqs. (\ref{cont}) and (\ref{ohm}) 
compose the basic set for the coupled dots, from which the excitation spectrum 
will be obtained.

In order to calculate the power absorption spectrum the same equation set 
has to be used but with the single replacement $\delta \varphi \to \delta \varphi +
\varphi^{ext}$, where the external field potential $\varphi^{ext}$ is added.
Then the frequency dependent power absorption spectrum is calculated in the standard way
using the following expression
\begin{equation}\label{power}
  P(\omega)=-\frac{\omega}{2\pi}\int_0^{2\pi/\omega} dt\,
  \sum_{j=1}^2 \int d^2 r \,{\bf j}_j \cdot
  \nabla \varphi^{ ext}.
\end{equation}

\section{Equilibrium properties}
\label{s3}

The first step in the static problem is obtained by replacing Poisson's
equation
(\ref{poisson}) by the homogeneous Laplace equation which is solved with 
the following boundary conditions for the potential in the dots
\begin{equation}\label{boundary}
  \left.\frac{\partial\varphi}{\partial z}\right|_{z=0}=\frac{2\pi e}{\epsilon}\rho_1,
  \quad
  \left.\frac{\partial\varphi}{\partial z}\right|_{z=a}=\frac{2\pi e}{\epsilon}\rho_2.
  \end{equation}
As was shown in Refs.~\onlinecite{ye94,nazin89,shikin91,tevosyan92} 
for the case of a single dot the most simple way
to treat this static problem is to use the oblate spheroidal coordinate system
centered in the dot, which is defined as follows:
$  x =\sqrt{(\sigma ^2+1)(1-\tau ^2)}\cos \theta , \;\;
  y =\sqrt{(\sigma ^2+1)(1-\tau ^2)}\sin \theta ,\;\; 
  z =\sigma \tau. $
%
In this coordinate system the disk corresponds to the region
$\sigma=0,\; 0\leq\tau\leq 1$. The advantage of this coordinate system is that
the variables in
Laplace's equation can be separated and the solutions can be expressed as
a linear superposition of the potential harmonics~\cite{morse}
$P_j^{|m|}(\tau)Q_j^{|m|}(i\sigma)$, with $j\geq|m|$. $P_j^{|m|}$ and
$Q_j^{|m|}$ are the associated  Legendre polynomials of the first and the second
kind, respectively. Now the static problem is reduced to the solution of
an algebraic set of equations for the expansion coefficients which follows from
the equilibrium conditions.
\par
We shall  use a similar approach in our coupled dot problem. 
In order
to have the most simple expressions for the main part of the electron-electron
interaction, namely the electron-electron interaction in the same dot, we introduce 
two coordinate systems each centered in the corresponding
dot. We rescale both coordinate systems, replacing $r \to rR_i$, and introduce 
the oblate spheroidal
coordinates in each of the dots, respectively. Next, we divide the potential
into two parts $\varphi=\varphi_1+\varphi_2$, each created by the electron
density in the corresponding dot, and introduce dimensionless variables by
performing the following transformations
\begin{equation}
  \rho _j(r) \longrightarrow \rho _j(r)\rho _j^{(0)}\sqrt{1-r^2},
  \quad \varphi _j \longrightarrow \frac{k_jR_j^2}{|e|}\varphi _j.
\end{equation}
By this transformation we explicitly took out the non-analyticity of the density 
function at the dot edge which is inherent in the case of a soft
confining potential.\cite{ye94}
If we choose the scaling factors $\rho _j^{(0)}=2\epsilon k_jR_j/\pi^2e^2$,
introduce the dimensionless numbers of electrons in the dots
$\ n_j=3N_j/2\pi R_j^2 \rho_j^{(0)}$, chemical potentials $v_j=\mu_j/k_jR_j^2$,
and ratios $\gamma =R_2/R_1$ and $\kappa =k_2/k_1$, we transform our static Eqs. (\ref{cond}),
(\ref{boundary}),
and (\ref{number}) into the following set of dimensionless equations
\begin{mathletters}
\begin{eqnarray}
\label{st1}
&&  r^2/2+\varphi_1^{dot}(r)+
  \gamma^2\kappa \varphi_2^{out}(\tilde{r}) = v_1,\\
\label{st2}
&&  r^2/2+\frac 1{\gamma ^2\kappa }\varphi_1^{out}(\tilde{r})
  +\varphi_2^{dot}(r) = v_2, \\
\label{st3}
&&  \left. \frac{\partial \varphi_j}{\partial \sigma}\right|_{\sigma=0}
   = \frac {2\tau^2}{\pi} \rho _j(r), \\
\label{st4}
&&  \int_0^1 dr r\sqrt{1-r^2}\rho_j(r) =
  \int_0^1 d\tau \tau^2 \rho_j(r) = \frac{n_j}{3}.
\end{eqnarray}
\end{mathletters}

\noindent 
Here the symbol $\tilde{r}$ stands for the coordinate in one dot
calculated in the coordinate system of the other dot. The superscripts
indicate whether the potential is created by the electron density in the same 
dot ($^{dot}$) or the electron density outside it ($^{out}$).

Laplace's equation must not be included into the set of equations because, as is
discussed above,
its solutions are already known \cite{ye94,morse} and can be presented as
\begin{equation}\label{pot_out}
\varphi_j^{out}(r)=\sum_l
  C_l^{(j)}P_{2l}(\tau)\frac{Q_{2l}(i\sigma)}{Q_{2l}(0)},
\end{equation}
where due to the circular symmetry of the equilibrium state only the even 
potential harmonics with $m=0$ have to be taken into account.
This general expression can be used for the potential outside the dot.
Inside the dot ($\sigma=0$) the potential can be presented in 
a more simple form
\begin{equation}\label{pot_dot}
\varphi_j^{dot}(r)=\sum_l
  C_l^{(j)}P_{2l}(\tau).
\end{equation}
The electron density in the dot can also be presented as an expansion in
Legendre polynomials
\begin{equation}\label{dens}
  \rho_j(r)=\frac{1}{\tau^2}\sum_l n_l^{(j)}P_{2l} (\tau).
\end{equation}
Here, the additional factor $\tau^2$ is included in order to simplify
the boundary condition (\ref{st3}) which now leads to the relation
$n_l^{(j)} = L_{2l}^0 C_l^{(j)}$ between the
coefficients in both expansions (\ref{pot_dot}) and (\ref{dens}),
with
\begin{equation}\label{gamma}
  L_l^{|m|}=\frac{\pi}{2}\frac{\Gamma\left((l+|m|)/2+1\right)
  \Gamma\left((l-|m|)/2+1\right)}{\Gamma\left((l+|m|+1)/2)\right)
  \Gamma\left((l-|m|+1)/2 \right)}
\end{equation}
where the symbol $\Gamma$ stands for the Gamma function.
Inserting expansions (\ref{pot_out}), (\ref{pot_dot}) and (\ref{dens}) into
Eqs.(\ref{st1}, \ref{st2}), multiplying them by $P_{2l}(\tau)$, integrating
over $\tau$ and using the Legendre polynomial orthogonality condition
\begin{equation}\label{ortho}
  \int_0^1 d\tau
  P_l^m(\tau)P_{l'}^m(\tau)=\frac{\delta_{l,l'}}{2l+1}\frac{(l+m)!}{(l- m)!},
\end{equation}
we obtain the following set of linear equations
\begin{mathletters}\label{equst}
\begin{eqnarray}
\label{equst1}
  C_l^{(1)}+ \kappa \sum_n T_{ln}(1/\gamma,d/\gamma) C_n^{(2)} &=&
  \left( v_1-\frac{1}{3}\right) \delta_{l,0}+\frac{1}{3}\delta_{l,2},  \\
\label{equst2}
  C_l^{(2)}+ \frac{1}{\kappa} \sum_n T_{ln}(\gamma,d) C_n^{(1)}&=&
  \left( v_2-\frac{1}{3}\right) \delta_{l,0}+\frac{1}{3}\delta_{l,2}.
\end{eqnarray}
\end{mathletters}
Here the symbol $T_{ln}$ stands for the overlap integral of the 
potential harmonics created by the electron densities  in the different dots
\begin{equation}
\label{def1}
  T_{ln}(\gamma,d) = \frac{(2l+1)}{\gamma^2} \int_0^1
  d\tau  P_{2l}(\tau) P_{2n}(\tilde{\tau})\frac{Q_{2n}(i\tilde{\sigma})}{Q_{2n}(0)}, 
\end{equation}
with $\tilde{\tau} = \left\{\left[
\sqrt{q^2+4d^2}-q\right]/2\right\}^{1/2},\;\;
\tilde{\sigma} = \left\{\left[ \sqrt{q^2+4d^2}+q\right]/2\right\}^{1/2}$ and 
$q = \gamma^2 (1-\tau^2)+d^2-1$.
In the above expressions we used the dimensionless distance between the two dots $d=a/R_1$.
Eqs. (\ref{equst}) should be supplemented by two more
conditions (as was shown in Ref.~\onlinecite{ye94})
\begin{equation}\label{condst}
  \sum_l L_{2l} P_{2l}(0)C_l^{(j)}=0,
\end{equation}
which guarantees the non singular behavior of the density functions
at the dot edge $\tau=0$.

Conditions (\ref{condst}) together with Eqs.(\ref{equst})
form a set of coupled linear algebraic equations for the coefficients
$v_j$ and $C_l^{(j)}$, which is equivalent to a nonhomogeneous matrix equation
for the vector $X=\{v_1,C_0^{(1)},\cdots,v_2,C_0^{(2)},\cdots\}$.
Only in the case of well separated dots ($a\to\infty$)
an analytical solution can be found. In that case we obtain
$v_1=v_2=n_1=n_2=1$
which corresponds to the single parabolic dot solution.~\cite{ye94}
For the general case we solved the problem numerically by means of
$QR$-decomposition of the truncated matrix.~\cite{recipes} We restricted
our consideration to 9 potential harmonics for each dot which was
sufficient to obtain an accuracy better that 0.1\%. 
We should note that in the case of different confining potentials $\kappa \ne 1$
or with an applied bias $eV \ne 0$ the numerical solution can not be found straightforwardly
because of the unknown ratio $\gamma$ of the dot radii.
We found $\gamma$ by solving the boundary
condition (\ref{bias}) iteratively.
After the ratio of the dot radii is found all quantities of interest can be
calculated just  using the scaling factors introduced in Sec.~\ref{s2}, namely,

\begin{eqnarray}
  N_1=\frac{n_1}{n_1+ \gamma^3\kappa n_2}N, \quad 
  N_2=\gamma^3\kappa\frac{n_2}{n_1}N_1, \nonumber \\
  R_1=\frac{1}{(n_1+\gamma^3\kappa n_2)^{(1/3)}}R_0, \quad R_2=\gamma R_1, \nonumber \\
  \rho_1^{(0)}=\frac{1}{(n_1+\gamma^3\kappa n_2)^{(1/3)}}\rho_0, \quad
  \rho_2^{(0)}=\gamma\kappa\rho_1^{(0)}. 
\end{eqnarray}
Here the symbol $N=N_1+N_2$ stands for the total number of electrons in boths dots, and
two scaling factors --- $R_0=(3\pi e^2N/4\epsilon  k_1)^{1/3}$ and $\rho_0 = 3N/2\pi R_0^2$
--- are used. The first one corresponds to the equilibrium radius of the first dot
when all electrons are situated in this dot. 
\par
The numerical results illustrating  the equilibrium electron densities in the dots are shown
in Figs.~\ref{figk1} and~\ref{figk2mu}. The density profiles for equivalent
coupled dots ($\kappa=1$) are given in Fig.~\ref{figk1} for various  
distances $a$ between the two dots.
In the case when there is no external bias ($\mu_1=\mu_2$) both dots have the same
density of electrons ($N_1=N_2=N/2$). In the limiting case of small distances
between the dots the result reduces to the single parabolic dot with density
profile $\rho_i \sim
\sqrt{1-r^2/R_0  ^2}$.
In the opposite case of large interdot distances we again have two parabolic 
dots where in
the above density functions the radius is scaled by $2^{1/3}$.
For intermediate distances, however, the density profile differs from that of a
single parabolic dot. This is illustrated in Fig.~\ref{figk1} by the thick
dashed-dotted curve which shows the parabolic dot density with the same dot size
and the same total number of electrons as those for the coupled dots with interdot
distance $a/R_0=0.3$.
The dependence of the dot size, $R_i$,  on the interdot separation is shown 
in the inset of Fig.~\ref{figk1}. With decreasing interdot separation the size
of the dots increases and the electron density profile is more spread out.
In Fig.~\ref{figk2mu} the density profiles for the case of different confining
potential strengths ($\kappa=2$, $\mu_1=\mu_2$) are given for
the same values of the interdot distances as in Fig.~\ref{figk1}. 
Notice that now the density profiles differ essentially from the
single parabolic dot solution. In the case of small interdot distances
even a   ring like distribution can be obtained when the electrons in the dot with 
the weakest confinement are pushed outside.
The above ring like distribution effect can be even more reinforced by applying
an external bias. Notice also that with decreasing interdot distance electrons
from dot 2 are moved into dot 1 in order to keep the chemical potential the same
(see inset of Fig.~\ref{figk2mu}(b)). The inset of Fig.~\ref{figk2mu}(a) shows
the dependence of the radius of the two dots on the distance between the two
dots.

\section{Magnetoplasma excitations}
\label{s4}

To solve the dynamic equations (\ref{poisson}), (\ref{cont}) and (\ref{ohm})
we use the same oblate spheroidal coordinate technique as in the static case
which was described in previous section. Restricting ourselves only to those modes
which can be excited in the dipole approximation we use the following
expansions for the harmonic deviations of the electron density
\begin{equation}\label{hdens}
  \delta\rho_j({\bf r})=\frac{1}{\tau^2}\sum_{l}g_{l}^{(j)}P_{2l+1}^{1}(\tau)
  e^{i\alpha\theta},
\label{dens2}
\end{equation}
the potential inside the dot
\begin{equation}\label{hpot_dot}
  \delta\varphi_j^{dot}({\bf r})=\sum_l c_{l}^{(j)}P_{2l+1}^{1}(\tau)e^{i\alpha\theta},
\end{equation}
and the potential outside the dot
\begin{equation}\label{hpot_out}
  \delta\varphi_j^{out}({\bf r})=\sum_l c_{l}^{(j)}P_{2l+1}^{1}(\tau)\frac{Q_{2l+1}^{1}(i\sigma)}
  {Q_{2l+1} ^{1}(0)}e^{i\alpha\theta}.
\end{equation}
The symbol $\alpha=\pm 1$ 
is the angular momentum of the excited harmonics
actually indicating the polarization of the circularly polarized electric field.
The above electron density and potential harmonics obey the same  boundary
conditions in the dots as given by Eq.~(\ref{st3}) which leads to the relation
$g_{l}^{(i)}=L_{2l+1}^{1}c_{l}^{(i)}$.
We will use the same dimensionless variables as in previous static case scaling
additionally the frequencies $\omega\to\omega\cdot\omega_1$ and
$\omega_c\to\omega_c\cdot\omega_1$. In those dimensionless
variables the continuity equation (\ref{cont}) becomes
\begin{mathletters}
\begin{eqnarray}
\label{dynamic1}
  \omega(\omega^2-\omega_c^2)\tau\delta\rho_1-\nabla\cdot\left(\tens{{s}}^{(1)}
  \cdot\nabla(\delta\varphi_1^{dot}+\delta\varphi_2^{out})\right)&=& 0, \\
\label{dynamic2}
  \frac{\omega}{\kappa}(\omega^2-\omega_c^2)\tau\delta
  \rho_2-\nabla\cdot\left(\tens{{s}}^{(2)}
  \cdot\nabla (\delta\varphi_1^{out}+\delta\varphi_2^{dot})\right)&=& 0,
\end{eqnarray}
\end{mathletters}
where for the sake of convenience we introduced the tensor $\tens{s}^{(i)}$,
which is 
proportional to the static densities, with the following components
\begin{equation}
s_{rr}^{(i)}=s_{\theta\theta}^{(i)}=\omega\tau\rho_i(\tau),\;\;
 s_{r\theta}^{(i)}=-s_{\theta r}^{(i)}=-i\omega_c\tau\rho_i(\tau).
\end{equation}
\par
Now inserting expansions (\ref{hdens}), (\ref{hpot_dot}) and (\ref{hpot_out}) into Eqs.
(\ref{dynamic1}, \ref{dynamic2}), multiplying the obtained expressions by
$\tau P_l^1(\tau)\exp(-i\alpha\theta)$, integrating over $\theta$ and $\tau$, and using
the orthogonality condition (\ref{ortho})
 we obtain the following set of equations for the coefficients $c_l^{(i)}$
\begin{mathletters}
\begin{eqnarray}
\label{vib1}
  \omega (\omega^2-\omega_c^2)
  \frac{2(l+1)(2l+1)}{4l+3}L_{2l+1}^1 c_l^{(1)}
  -\sum_n V^{(1)}_{ln} c_n^{(1)}- \kappa\sum_n
  W^{(2)}_{ln}(1/\gamma,d/\gamma)c_n^{(2)} &=& 0 ,\\
\label{vib2}
  \frac{\omega}{\kappa} (\omega^2-\omega_c^2)
  \frac{2(l+1)(2l+1)}{4l+3}L_{2l+1}^1 c_l^{(2)}-
  \frac{1}{\kappa}\sum_n W_{ln}^{(1}(\gamma,d)c_n^{(1)}- \sum_n
  V^{(2)}_{ln}c_n^{(2)} &=& 0,
\end{eqnarray}
\end{mathletters}
which is cast into a homogeneous matrix equation
\begin{equation}
  {\cal M}(\omega)X=0,
\end{equation}
for the vector $X=\{c_0^{(1)},c_1^{(1)},\cdots,c_0^{(2)},c_1^{(2)},\cdots\}$.
The dynamic matrix ${\cal M}$ is composed of the overlap integrals of the potential
components belonging to the same dot
\begin{eqnarray}\label{mediag}
  V_{ln}^{(i)}
  &=& \frac{1}{2\pi}\int_0^{2\pi}d\theta \int_0^1 d\tau \tau P_{2l+1}^1(\tau)
  e^{-i\alpha\theta} \nabla\cdot \left(\tens{s}^{(i)}\cdot \nabla
  P_{2n+1}^1(\tau)e^{i\alpha\theta}\right) \nonumber \\
  &=& \int_0^1 d\tau P_{2l+1}^1(\tau) \left\{ -\frac{s_{rr}^{(i)}}{\tau}[n(n+1)-1]
  \right. \nonumber \\
&&  \left. +(1-\tau^2) \left[ \frac{d}
  {d \tau} \left( \frac{s_{rr}^{(i)}}{\tau} \right) \right] \frac{\partial}
  {\partial \tau} -i\alpha \left(  \frac{ds_{r\theta}^{(i)}}
  {d\tau} \right) \right\} P_{2n+1}^1(\tau).
\end{eqnarray}
and to the different dots
\begin{eqnarray}
  \lefteqn{W_{ln}^{(i)}(\gamma,d) =
  \int_0^1 d\tau P_{2n+1}^1(\tilde{\tau})\frac{Q_{2n+1}^1(i\tilde{\sigma})}{Q_{2n+1}^1(0)}
  \left\{ -\frac{ s_{rr}^{(i)}}{\tau}[l(l+1)-1] \right.}\nonumber \\
 & &\left. +(1-\tau^2) \left[ \frac{d}
{d \tau} \left( \frac{s_{rr}^{(i)}}{\tau} \right) \right] \frac{\partial}
{\partial \tau} + \left(  \frac{ds_{r\theta}^{(i)}}
{d\tau} \right) \right\} P_{2l+1}^1(\tau). \\
\end{eqnarray}
The functions $\tilde{\tau}, \tilde{\sigma}$ and $q$ are defined as in the
stationary case. 

The magnetoplasma excitation frequencies $\omega_j$ are obtained as solutions of the
following equation
\begin{equation}\label{haha}
  \mbox{det}{\cal M}(\omega) = 0.
\end{equation}
The matrix determinant was calculated numerically by means of  LU-decomposition \cite{recipes}
truncating the matrix up to $7$ potential components for each dot which
guaranteed the same accuracy as in the equilibrium case.

\section{Power absorption}
\label{s5}

We restrict our consideration to the case of weak dissipation when the
magnetoplasma
excitation spectrum can be most easily revealed. In that particular case
the power absorption can be calculated using the same set of dynamic 
equations (\ref{dynamic1}, \ref{dynamic2}) but now with the external electric field added.
Our system is circular symmetric and therefore it is most convenient to consider 
the circularly polarized electric field
${\bf E}^{\mbox{ext}}=E_0 ({\bf e}_x+i\alpha{\bf e}_y)\exp(i\omega t)$
with the corresponding external potential amplitude $\varphi^{\mbox{ext}}
({\bf r})=-E_0(x+i\alpha y)$.
Now inserting the external potential into the set of dynamic equations
(\ref{dynamic1}, \ref{dynamic2}), performing the same procedure as before,
and scaling the vector $X \to X \sqrt{2\pi} |e|E_0/k_1 R_1$ 
we arrive at the following nonhomogeneous matrix
equation
\begin{equation}\label{nonhom}
  {\cal M}(\tilde{\omega})X = B.
\end{equation}
Here, the vector $B=\{b_0^{(1)},b_1^{(1)},\cdots,b_0^{(2)},b_1^{(2)},\cdots\}$ is
composed of the following components
\begin{equation}
  b_l^{(i)} = 2(l+1)(2l+1)(\omega+\alpha\omega_c) \sum_{l'}S_{l',l} n_{l'}^{(i)}
  \cdot \left\{ \begin{array}{cc} 1, & i=1, \\ 1/\kappa\gamma, & i=2, \end{array}
  \right.
\end{equation}
and
\begin{equation}
  S_{l',l} = \int_0^1 \frac{d\tau}   {\tau} P_{2l'}(\tau)P_{2l+1}(\tau)
\end{equation}
is an overlap integral of even  and odd Legendre polynomials.
\par
Besides, in order to take some small dissipation into account we applied the standard
procedure making the following frequency replacement
$\omega\to\tilde{\omega}=\omega+i\eta$. The symbol $\eta$ represents a phenomenological
inverse electron relaxation time, and takes into account the internal dissipation
within the electron fluid in the dot.
According to expression (\ref{power}) the power absorption
can be presented as
\begin{equation}
  P(\omega)=-\omega P_0 \mbox{Im}L_1^1\{c_0^{(1)}+
  \kappa\gamma^4c_0^{(2)}\},
\end{equation}
where we have introduced the absorption power scaling factor
$P_0=4\epsilon E_0^2 R_1^3 \omega_1 /3\pi $.
Formally, the dimensionless absorption power can be expressed in
matrix form
\begin{equation}\label{pw}
  p(\omega)=P(\omega)/P_0=-\mbox{Im}\,\{ T\cdot X \},
\end{equation}
where the vector $T=\{\omega L_1^1,0,\cdots,0,\omega\kappa\gamma^4 L_1^1,0,\cdots,0 \}$
has only two non zero components.

Equation (\ref{nonhom}) and the power absorption (\ref{pw}) was solved numerically
using the $SVD$ (singular value decomposition) \cite{recipes} of the dynamical matrix
\begin{equation}\label{svd}
  {\cal M}(\tilde{\omega}) = {\cal U}\Lambda(\tilde{\omega}){\cal V}^T,
\end{equation}
where ${\cal U}$ and ${\cal V}$ are unitary matrices (superscript $^T$ stands for
transposed matrix) and $\Lambda$ is the diagonal matrix composed of the eigenvalues
$\lambda_i(\tilde{\omega})$ of the dynamical matrix ${\cal M}$. Due to the decomposition
(\ref{svd}) the solution of equation (\ref{nonhom}) can be presented as
\begin{equation}
  X = {\cal V}\Lambda^{-1}(\tilde{\omega}){\cal U}^T B.
\end{equation}
Inserting it into expression (\ref{pw}) we obtain the general expression for the
power absorption
\begin{equation}\label{gpw}
  p(\omega)=-\mbox{Im}\,\{ T\cdot {\cal V}\Lambda^{-1}(\tilde{\omega}){\cal U}^T B\}.
\end{equation}

This expression can be further simplified taking into account that we are considering
the case of weak dissipation ($\eta \to 0$) only. In that case the power absorption
consists of narrow Lorentzians located at the magnetoplasma excitation
frequencies $\omega_j$. According to Eq.~(\ref{haha}) the corresponding conditions
$\lambda_i(\omega_i)=0$ should be fulfilled. This enables
us to use the following expansion
\begin{equation}
  \lambda_i(\tilde{\omega}) = \lambda'_i(\omega_i)(\omega-\omega_i+i\eta).
\end{equation}
Inserting it into expression (\ref{gpw}) gives the following final expression
for the power absorption
\begin{eqnarray}
  p(\omega) &=& -\mbox{Im}\,\sum_{i}\frac{(T\cdot V_i)(U_i^T\cdot B)}
  {\lambda'_i(\omega_i)(\omega-\omega_i+i\eta)} = \sum_{i}\frac{\eta f_i}
  {(\omega-\omega_i)^2+\eta^2},
\end{eqnarray}
where we introduced the coefficients
\begin{equation}\label{oscstr}
  f_i = \mbox{Re}\,\frac{(T\cdot V_i)(U_i^T\cdot B)}{\lambda'_i(\omega_i)},
\end{equation}
and the symbol $V_i$ stands for the $i$-th vector-column of matrix ${\cal V}$, and
the symbol $U_i^T$ for the $i$-th vector-row of the transposed matrix  ${\cal U}^T$.
Note that we neglected the imaginary part of the right hand
side of expression~(\ref{oscstr}), because for weak 
dissipation it only leads to a negligible
eigenfrequency shift. For the same reason the other quantities
(like ${\cal U}$, ${\cal V}$, $\lambda'$) should be estimated at
$\omega=\omega_i$.

In the case of weak dissipation the quantities (\ref{oscstr}) together
with the mode frequencies $\omega_i$ characterise the power absorption.
By analogy to quantum mechanical problems they can be referred to as the classical
oscillator strengths. As is shown in Appendix~\ref{a1} these oscillator
strengths obey a general
sum rule like their quantum mechanical counterparts, indicating that the interaction
variation in the system leads to the redistribution of power absorption among
various vibration modes while the total power absorption remains unchanged.
 
\section{Results and discussion}
\label{s6}

For the case of equivalent dots $\kappa=1$ and $\mu_1=\mu_2$ the eigenfrequencies and oscillator strengths
are given in Fig.~\ref{figfreqk1}.
The frequencies in the limiting case of $a\to\infty$ are shown in
Fig.~\ref{figfreqk1}(a) as function of the applied magnetic field
($\omega_c$).
The eigenfrequencies for clockwise
circular polarization ($\alpha=1$) are shown by solid curves while the dashed
curves indicate the
eigenfrequencies for the opposite polarization ($\alpha=-1$).
In the limit case $a\rightarrow\infty$ of decoupled dots one
finds that 
$W_{ln}^{(i)}=0$ and $V_{ln}^{(i)}$ is a diagonal matrix (see Eqs. (4.9) and
(4.10)). Then the eigenmodes
are the basisfunctions $P_{2l+1}^1(\tau)$ and the spectrum branches can be uniquely 
labeled with the numbers $(l,\alpha=\pm 1)$.
In this limit the eigenfrequencies are twofold degenerate and coincide
with the single parabolic dot frequencies presented in
Ref.~\onlinecite{ye94}.
In the low frequency region a large set of branches are found which correspond 
to the edge modes. They are
not resolved and shown up as a shaded area. In Fig.~\ref{figfreqk1}(b) the
spectrum for an interdot separation of $a/R_0=0.5$ is presented. 
For the sake of convenience we still label the branches by the previous 
$l$-values in order
to show the correspondance of the branch to its counterpart in the $a\to\infty$  
limiting case. In this case in fact several $l$-values are mixed.
Note that 
when the distance  between the layers  decreases
the degeneracy is lifted and all branches
are split into two and are shifted down in frequency except for the two branches
depicted by the thick
curves which correspond
to the center of mass motion of the whole system of electrons in both dots. 
Only the latter two frequencies have non zero
oscillator strengths which do not depend on the interdot
distance and are shown in Fig.~\ref{figfreqk1}(c). This is in accordance with Kohn's theorem~\cite{kohn} which can
be easily generalized
to our vertically aligned two dot system for the case of identical dots, i.e.
$\kappa=1$. The decrease of the eigenfrequencies
with decreasing interdot distance is a consequence of the increase of
the repulsive interaction of the electrons in the different dots and is discussed in more detail
in Appendix~\ref{a2}.

The case of nonequivalent dots is illustrated in Figs.~\ref{figfreqk2mu} and 
\ref{figfreqk2d05} where the results
for $\kappa=2$ and $\mu_1=\mu_2$ are presented. The frequencies for the
$a\to\infty$ limit
are shown in Fig.~\ref{figfreqk2mu}(a) , and the corresponding oscillator strengths are
shown in Fig.~\ref{figfreqk2mu}(b). In this case the system consists of two noninteracting
parabolic dots where the frequencies are scaled by the factor $\sqrt{\kappa}=
\sqrt{2}$. The branches are labeled with the corresponding $l$-value. 
The non-zero oscillator strenghts are  those which corresponds to the
center of mass motions of the electron systems in the separate dots. Note that the
oscillator strengths obey the sum rule that the sum of the oscillator strengths
equals 1 independent of $\omega_c$.
The redistribution of the oscillator strengths between the separate dot branches
actually reflects the electron number redistribution between the dots (see
Fig.~\ref{figk2mu}).
It is also evident that those spectrum branches which are closer  to the cyclotron
resonance branch ($\omega=\omega_c$) have essentially larger oscillator strengths. 

The spectrum branches and the oscillator strengths change fundamentally when
nonequivalent dots are put closer to each other as is shown in Fig.~\ref{figfreqk2d05}.
Now all the branches are shifted down in frequency and non-zero oscillator strengths corresponding
to other then the center of mass motion branches start to appear. For instance,
in Fig.~\ref{figfreqk2d05}(b)
the oscillator strengths corresponding to the frequency branches ($l=1$) are
non-zero as shown by the curves labeled 5 and 6.
The reason is that for nonequivalent dots the electron density
distributions are far away from the parabolic dot distribution (see, Fig.~\ref{figk2mu}),
and Kohn's theorem is no longer valid.
Note also from Fig.~\ref{figfreqk2d05}(a) that the frequency
shift to lower frequencies is much more pronounced as compared to the case of
equivalent dots (Fig.~\ref{figfreqk1}(b)).
The above frequency shift can be even more amplified when an external
bias is applied. This is illustrated in Fig.~\ref{figk2mu2} where the results for
$\kappa=2$ and $\mu_1-\mu_2=-0.2$ are given. From
Fig.~\ref{figk2mu2}(a) we see that now the $(l=1)$ branches start to interfere with the 
$(l=0)$ ones. Although there is still no intersection with the branches of the same
polarization the branch ($l=1$, $\alpha=1$) approaches the cyclotron resonance
line ($\omega=\omega_c$) so closely that the oscillator strength
(see Fig.~\ref{figk2mu2}(b)) is transferred from the main 
branch ($l=0$, $\alpha=1$).

We would like to note that in the above case of nonequivalent dots when the
system no longer obeys Kohn's theorem the edge modes
have small but non-zero oscillator strengths. In order to reveal those
weak modes we performed a more accurate dynamic matrix diagonalisation for the
case $\mu_1=\mu_2$, $a/R_0=0.5$ and $\kappa=2$ by restricting ourselves to 
the $l=0$ and $l=1$ components for each dot. 
The results for the edge modes (curves 9 and 10) together with the oscillator
strengths for the upper $l=1$ modes (curves 7 and 8) are given in
Fig.~\ref{figedge}. We see that the oscillator strengths become very small for
spectrum branches which are substantially different from the above mentioned $\omega=\omega_c$.

\section{Conclusions}
\label{s7}
In this paper, we have studied the magnetoplasma excitations in two vertically
coupled electron dots in the dipole approximation. We used the classical
hydrodynamic approach which is
adequate in the limit of a large number of electrons. The confinement potential
was taken parabolic in both dots. We calculated the equilibrium electron 
density distribution as function of the interdot distance. As function of the
magnetic field the frequency spectrum and the corresponding oscillator strengths
are calculated. 
\par
In the case of equivalent dots, i.e. both dots have the same parabolic
confinement strength, the equilibrium density distribution is similar to the
single dot
parabolic one, and due to the generalized Kohn's theorem only the center of mass
motion can be excited. 
\par
If both dots have a different parabolic confinement the electron densities can
differ appreciably from the parabolic case. Even ring like electron distributions are found.
Due to the fact that both dots have different confinement potentials Kohn's
theorem is no longer valid and other modes
then the center of mass motions can be excited. The largest oscillator
strengths are found for the modes closest to the cyclotron frequency. If an
external bias potential is applied the oscillator strengths of other then the
center of mass modes are even more enhanced. Crossover phenomena can be
observed in this case. For nonequivalent dots, even the edge modes have small
but non-zero oscillator strengths.

\section{Acknowledgments}

B.P. is an Aspirant and F.M.P. a Research Director of the
Flemish Science Foundation (FWO-Vlaanderen). This work is partially supported by a
NATO-linkage
Collaborative Research Grant, the FWO-Vlaanderen and the `Interuniversity Poles
of Attraction Program - Belgian State, Prime Minister's Office - Federal Office
for Scientific, Technical and Cultural Affairs'.

\appendix

\section{Sum rule for the classical oscillator strengths}
\label{a1}
In this Appendix we present the definition of the classical oscillator strength
for a general linear mechanical system with weak dissipation and we
proof a sum rule for these oscillator strengths.
We consider a mechanical system interacting with an
external force which can be described by the following matrix
equation
\begin{equation}\label{mechsyst}
  \left( \frac{d}{dt}-{\cal L} \right)X(t) = A\, f(t).
\end{equation}
Here symbol $X$ stands for the vector composed of the system variables: coordinates,
velocities {\it {etc.}} ${\cal L}$ is the dynamical matrix describing the mechanical
system, $A$ is the unit length vector ($A^2=1$) characterizing the
application of the external force, and the scalar function $f(t)$ stands for the
external force itself. The power absorption is usually calculated for a
periodic driving force
\begin{equation}\label{periodic}
  f(t) = f(\omega)\cos(\omega t) = \mbox{Re}\left\{fe^{i\omega t}\right\}.
\end{equation}
In that case the averaged power absorption is
\begin{eqnarray}\label{power1}
  P(\omega) &=& \frac{\omega}{2\pi}\int_0^{2\pi/\omega}dt\, P(t)
  = \frac{\omega}{2\pi}\int_0^{2\pi/\omega}dt\,f(t)(A^T\cdot X(t))
  \nonumber \\
  &=& \frac{1}{2} \mbox{Re} \left\{f^{*}(\omega) (A^T \cdot X(\omega)) \right\},
\end{eqnarray}
where the coordinate vector Fourier component obeys the following equation
\begin{equation}\label{harm}
  \left( i\omega-{\cal L} \right)X(\omega) = A\, f.
\end{equation}
Note that our set of dynamic equations (\ref{vib1}, \ref{vib2}) can be easily
reduced to this type
of equation by just introducing additional variables $Y=\omega X$, $Z=\omega Y$ and constructing the
resulting vector $\hat{X}=\{X,Y,Z\}$ of all those variables.
\par
Now we use the following dynamical matrix representation
\begin{equation}\label{decomp}
  {\cal L} = {\cal U}{\cal D}{\cal U}^{-1},
\end{equation}
where ${\cal D}$ is the diagonal matrix composed of the eigenvalues
$i\omega_i$
corresponding to the system eigenfrequencies. Using that expansion and
the substitution $\omega \to \tilde{\omega}=\omega+i\eta$ we obtain
the solution of equation (\ref{harm})
\begin{eqnarray}\label{sol}
  X(\omega) &=& {\cal U}\left( i\tilde{\omega}-{\cal D} \right)^{-1}
  {\cal U}^{-1} A\, f \nonumber \\
  &=& f\sum_{i=1}^{n}U_i \{i(\omega-\omega_i)-\eta\}^{-1}(U_i^{-1}\cdot A),
\end{eqnarray}
and the expression for the power absorption
\begin{eqnarray}\label{powom}
  P(\omega) &=& \frac{1}{2}f^2 \mbox{Re}\sum_{i=1}^{n}(A^T \cdot U_i)
  \{i(\omega-\omega_i)-\eta\}^{-1}(U_i^{-1}\cdot A) \nonumber \\
  &=& \frac{1}{2}f^2 \sum_{i=1}^{n} \frac{f_i \eta}{(\omega-\omega_i)^2
  +\eta^2},
\end{eqnarray}
where
\begin{equation}\label{osc}
  f_i = \mbox{Re} \left\{(A^T \cdot U_i)(U_i^{-1}\cdot A)\right\},
\end{equation}
is the oscillator strength. Like in Sec.~\ref{s5} symbol $U_i$ stands for the $i$-th
vector-column of matrix ${\cal U}$, and the
symbol $U_i^{-1}$ is the $i$-th vector-row of matrix ${\cal U}^{-1}$, and
the term $\mbox{Im} \left\{(A^T \cdot U_i)(U_i^{-1}\cdot A)\right\}$
is neglected because in the case of weak dissipation it gives a negligible
eigenfrequency shift.

The oscillator strengths obey a useful sum rule which follows straightforwardly
from definition (\ref{osc}), namely,
\begin{eqnarray}\label{sumrule}
  \sum_{i=1}^n f_i &=& \mbox{Re} \sum_{i=1}^n \left\{(A^T \cdot U_i)
  (U_i^{-1}\cdot A)\right\} \nonumber \\
  &=& \mbox{Re} (A^T \cdot UU^{-1} \cdot A) = (A^T \cdot A)=1.
\end{eqnarray}
We obtained a relation which shows that the sum of the oscillator strengths
does not
depend on the properties of the dynamical system. 
Note we omitted symbol $\mbox{Re}$ in the last equation (\ref{sumrule})
because the classical equations of type (\ref{mechsyst}) are always formulated
with real coefficients.

The obtained sum rule has a very simple physical meaning. It is just related
to the total power absorption in the case of the impact external force
$f(t) = A \delta(t)$.
Actually, in that case calculating the total power absorption
\begin{equation}
  P = \int_{-\infty}^{\infty} dt f(t) (A^T \cdot X(t))
     = \int_{-\infty}^{\infty} dt \delta(t) (A^T \cdot X(t)),
\end{equation}
we only need to know the solution in the limit of small time values.
That solution can be obtained just neglecting the dynamical matrix ${\cal L}$
in equation (A1) as compared with the time derivative. So, the solution reads
\begin{equation}
  X(t) = A \int_{-\infty}^t dt \delta(t).
\end{equation}
and the total power absorption 
\begin{equation}
  P = \int_{-\infty}^{\infty} dt f(t) (A^T \cdot X(t)) = (A^T \cdot A)
  \int_{-\infty}^{\infty} dt \delta(t) \int_{-\infty}^t dt' \delta(t') =
  \frac{1}{2}\sum_{i=1}^n f_i,
\end{equation}
coincides with half of the above defined sum. Thus, the total absorption power
does not depend on the properties of the dynamical system, and that confirms
that the sum of oscillator strengths is a universal quantity.

\section{Two coupled oscillators with repulsive interaction}
\label{a2}

In Sec. VI it was shown that all the spectrum branches show a general behavior,
namely the frequencies decrease with decreasing
interdot distance. Such behavior is caused by the repulsive interaction
of the electrons belonging to different dots, and can be explained using the
following simple model. We focus our attention to the
center of mass motions of the electrons of the seperate dots.
In the case the dots are decoupled ($a\rightarrow\infty$) those motions
can be treated as two independent oscillators in a magnetic field
perpendicular to the plane of motion. When the dots are put closer
to each other, say, to some distance $a \gg R_i$, the interdot electron
interaction can be approximated by
\begin{equation}
  \frac{e^2}{\sqrt{({\bf r}_1-{\bf r}_2)^2+a^2}} \approx \frac{e^2}{a}
  -\frac{e^2}{2a^3}({\bf r}_1-{\bf r}_2)^2.
\end{equation}
Here the coordinate ${\bf r}_i$ corresponds to the lateral center of mass
motion of the 
electrons in dot $i$. The first term in the above expression gives the
total energy shift while the second one should be inserted into the equations
describing the system oscillations. Thus our model system consists of
two harmonic oscillators with a repulsive harmonic interaction in a magnetic
field. If also an external clockwise circularly polarized electric field is
applied, the equations read
\begin{mathletters}
\begin{eqnarray}
  m\ddot{{\bf r}}_1+k_1{\bf r}_1+\frac{e}{c}\dot{{\bf r}}_1\times {\bf B} -
  \frac{e^2}{a^3}({\bf r}_1-{\bf r}_2) &=& eN_1 E_0 (\cos (\omega t) {\bf e_x}
  -\sin(\omega t) {\bf e_y}),\\
  m\ddot{{\bf r}}_2+k_2{\bf r}_1+\frac{e}{c}\dot{{\bf r}}_2\times {\bf B} +
  \frac{e^2}{a^3}({\bf r}_1-{\bf r}_2) &=& eN_2 E_0 (\cos (\omega t) {\bf e_x}
  -\sin(\omega t) {\bf e_y}),
\end{eqnarray}
\end{mathletters}
where $n_1$ and $N_2$ are the total number of electrons in each dot, and $E_0$
is the amplitude of the applied electric field. 
This linear set of equations can be solved by introducing the complex
variables $z_i=x_i+iy_i$ which leads to the following set of equations
\begin{eqnarray}
  \ddot{z}_1 +i\omega_c \dot{z}_1 + \omega_1^2 z_1 -\Omega_0^2(z_1-z_2) &=&
  \frac{eN_1E_0}{m}e^{-i\omega t},
  \nonumber \\
  \ddot{z}_2 +i\omega_c \dot{z}_2 + \omega_2^2 z_2 +\Omega_0^2(z_1-z_2) &=&
  \frac{eN_2E_0}{m}e^{-i\omega t}.
\end{eqnarray}
Here $\Omega_0^2=e^2/ma^3$.
Now inserting $z_i \sim \exp(-i\omega t)$ into the homogenious version of the 
above expressions and mapping
the four obtained eigenfrequencies for both polarizations onto the positive
frequencies we have
\begin{equation}
  \omega^{1,2,3,4} = 
  \sqrt{\left(\frac{\omega_c}{2}\right)^2+\frac{\omega_1^2+\omega_2^2}{2}-\Omega_0^2
  \pm \sqrt{\left( \frac{\omega_1^2-\omega_2^2}{2}\right)^2+\Omega_0^4}} \pm
  \frac{\omega_c}{2}.
\end{equation}
It is easily seen that the derivative
\begin{equation}
  \left.\frac{d\omega^{1,2,3,4}}{d\Omega_0^2}\right|_{\Omega_0=0} =
  \frac{-1}{\sqrt{\left(\frac{\omega_c}{2}\right)^2+\frac{\omega_1^2+\omega_2^2}{2}
  \pm \left( \frac{\omega_1^2-\omega_2^2}{2}\right)}},
\end{equation}
is always negative what confirms that the interdot interaction of electrons
causes
a negative frequency shift, and thus, illustrates the previously obtained
negative
shift of the spectrum branches with decreasing interdot distance. 
\par
The power
absorption for this model system can also easily be calculated. It reads
\begin{equation}
P(\omega)=\frac{1}{2m}e(N_1+N_2)E_0\sum_{i=1}^4\frac{f_i\eta}{(\omega-\omega_i)^2+\eta^2},
\end{equation}
where the oscillator strengths are given by
\begin{equation}
f_i=\frac{(\omega^i)^3+\omega_c(\omega^i)^2+
\left(\Omega_0^2-\frac{N_1}{N_1+N_2}
(\omega_2^2-\Omega_0^2)-\frac{N_2}{N_1+N_2}(\omega_1^2-\Omega_0^2)\right)\omega^i} 
{\prod_{j\ne i}(\omega^i-\omega^j)}.
\end{equation}
It can be checked straightforwardly that
the oscillator strengths obey the sum rule $\sum_{i=1}^4 f_i=1$. They show the
correct qualitative behavior as function of the cyclotron frequency as shown
for the coupled dots in Sect.~\ref{s6}.

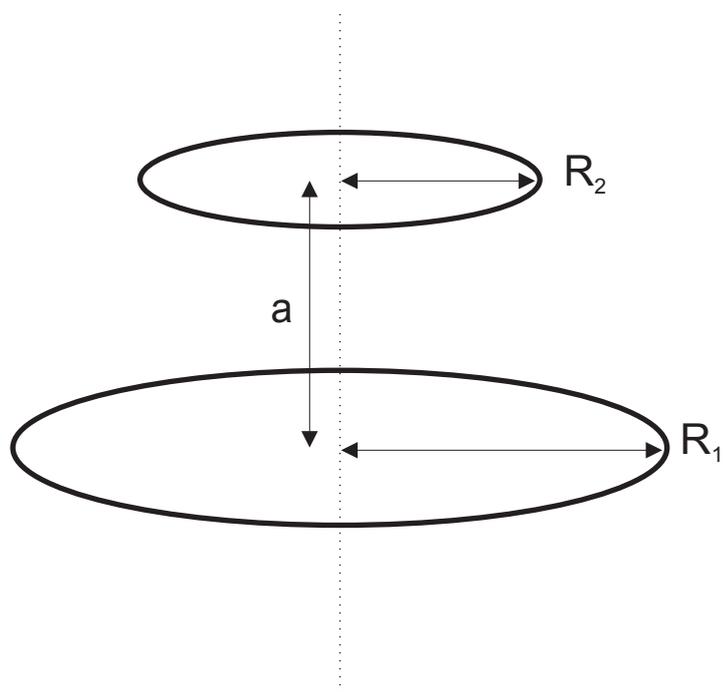
\begin{figure}
\caption{The layout of the structure.}
\label{layout}
\end{figure}
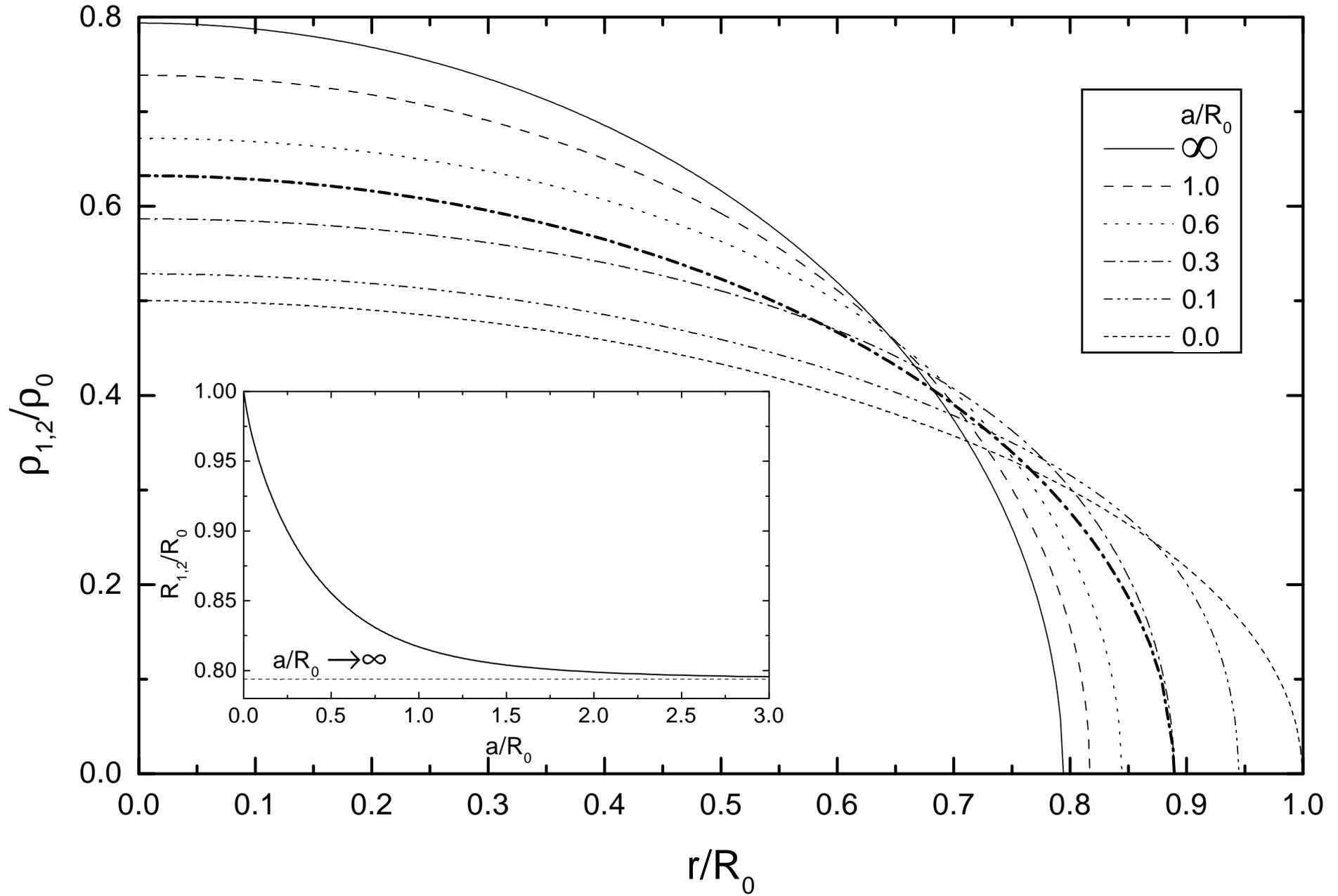
\begin{figure}
\caption{The equilibrium electron density profiles for the system with
$\kappa=1$ and $\mu_1=\mu_2$ for different values of $a/R_0$.
The inset shows the radius of the dots
as function of $a/R_0$.}
\label{figk1}
\end{figure}
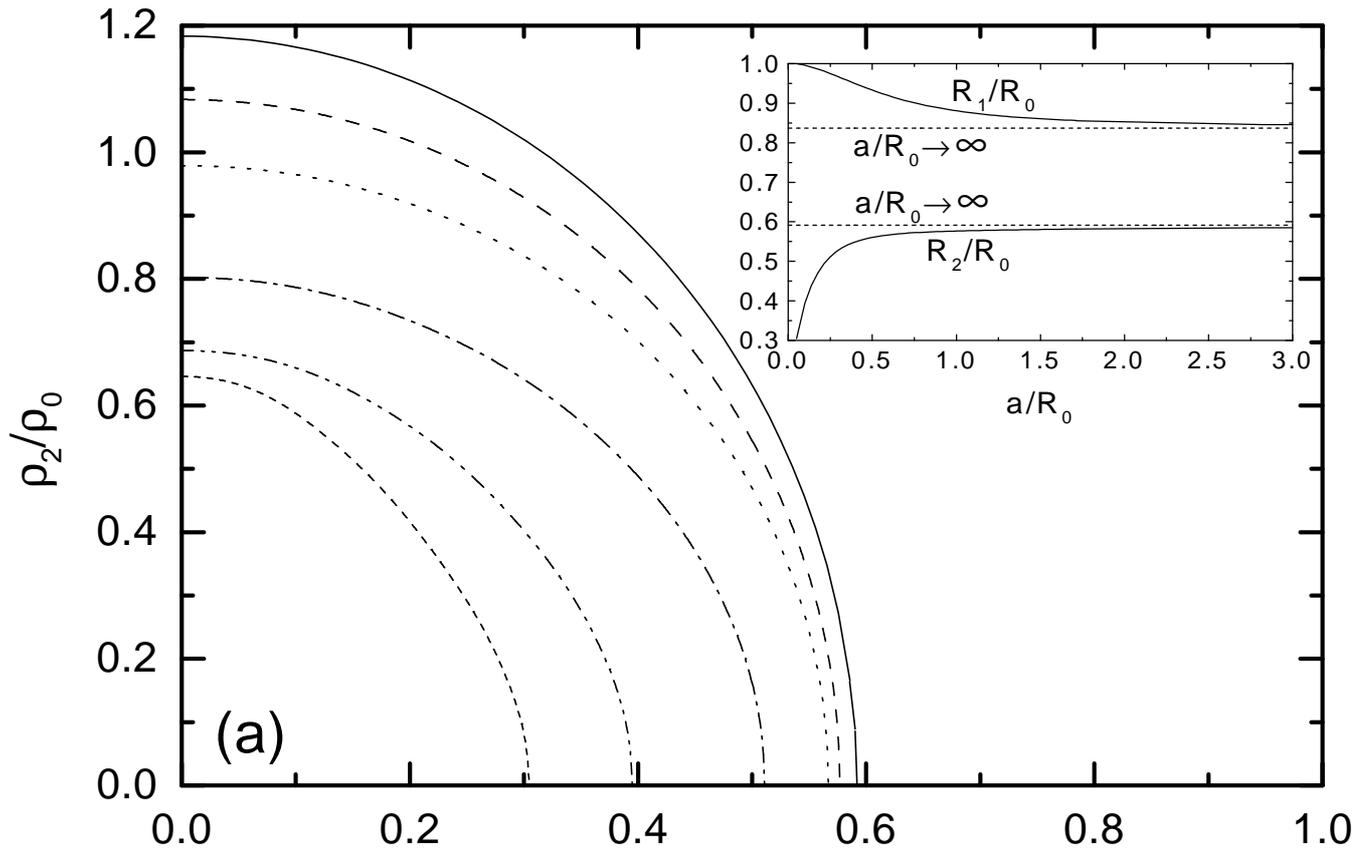
\begin{figure}
\caption{The equilibrium electron density profiles for the system with
$\kappa=2$ and $\mu_1=\mu_2$ for the same values of $a/R_0$ as in
Fig.~\ref{figk1}. 
The inset in (a) shows the radii 
of the dots, the inset in (b) the number of electrons in dot 1 
as function of $a/R_0$.}
\label{figk2mu}
\end{figure}
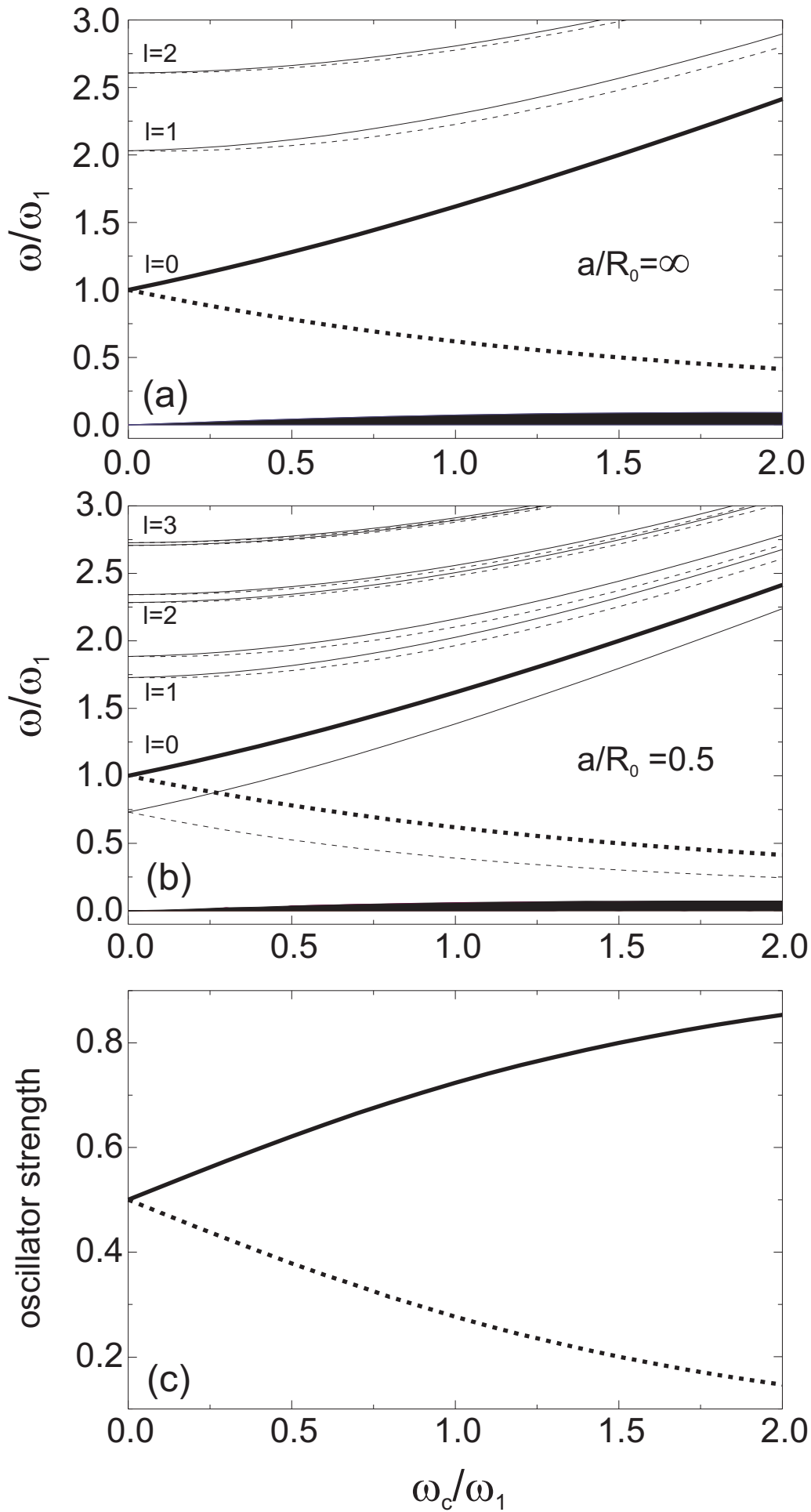
\begin{figure}
\caption{The frequency spectrum for the system with
$\kappa=1$. The limiting case $a/R_0=\infty$ is shown in (a) where the
branches are labeled by the corresponding $l$ value. The case
$a/R_0$=0.5
is shown in (b). 
The solid lines correspond to the $(\alpha=1)$ branches, the dotted lines to the $(\alpha=-1)$
ones.  
The branches that can be excited in the
dipole approximation are indicated by thick curves. The corresponding oscillator strengths are shown in (c).}
\label{figfreqk1}
\end{figure}
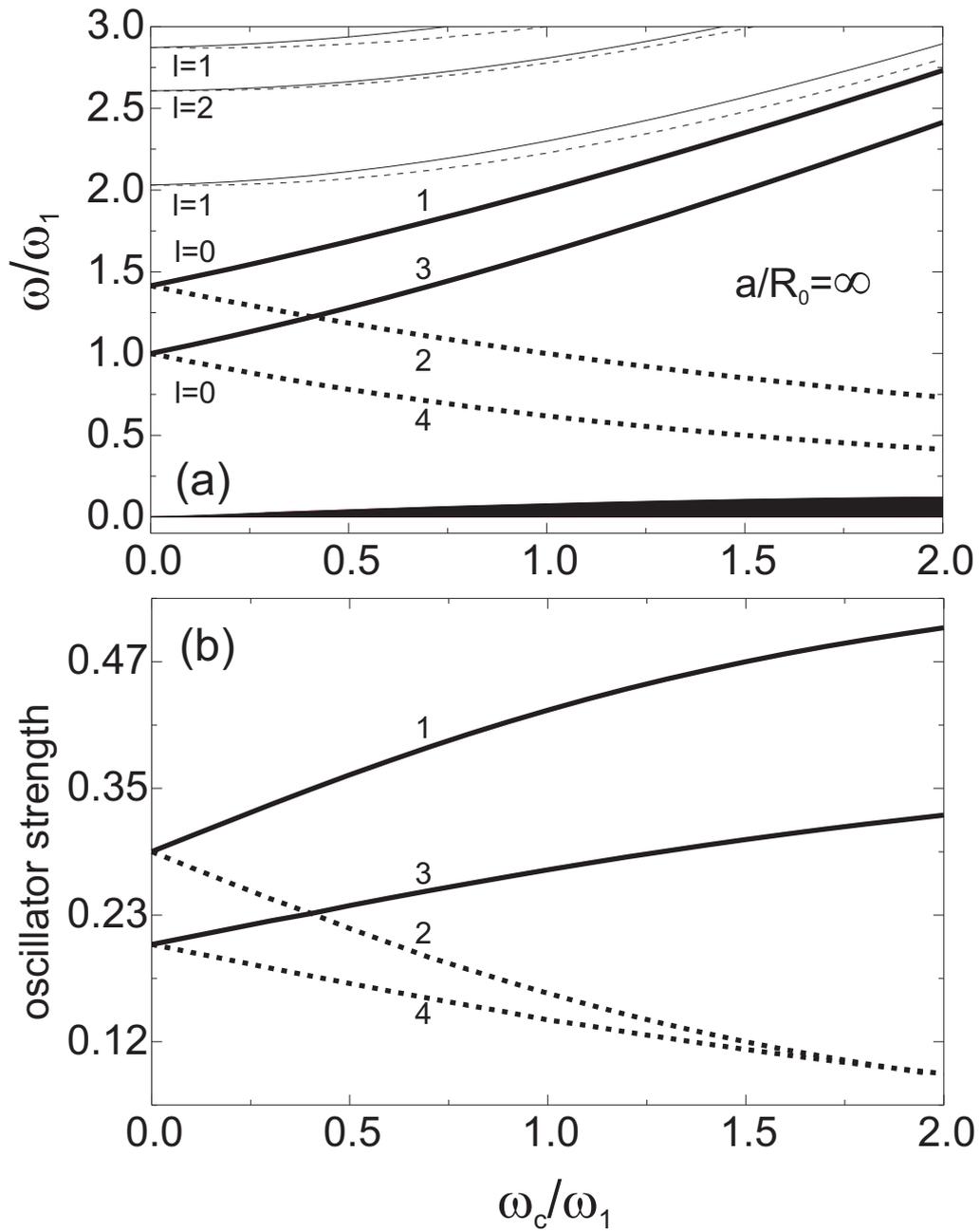
\begin{figure}
\caption{The frequency spectrum and the oscillator strength for the system with 
$\kappa=2$ and $\mu_1=\mu_2$ in the limit $a/R_0=\infty$. The notation
is the same as in Fig.~\ref{figfreqk1}. The correspondence between the spectrum branches
and the oscillator strengths are indicated by the numbers 1-4.}
\label{figfreqk2mu}
\end{figure}
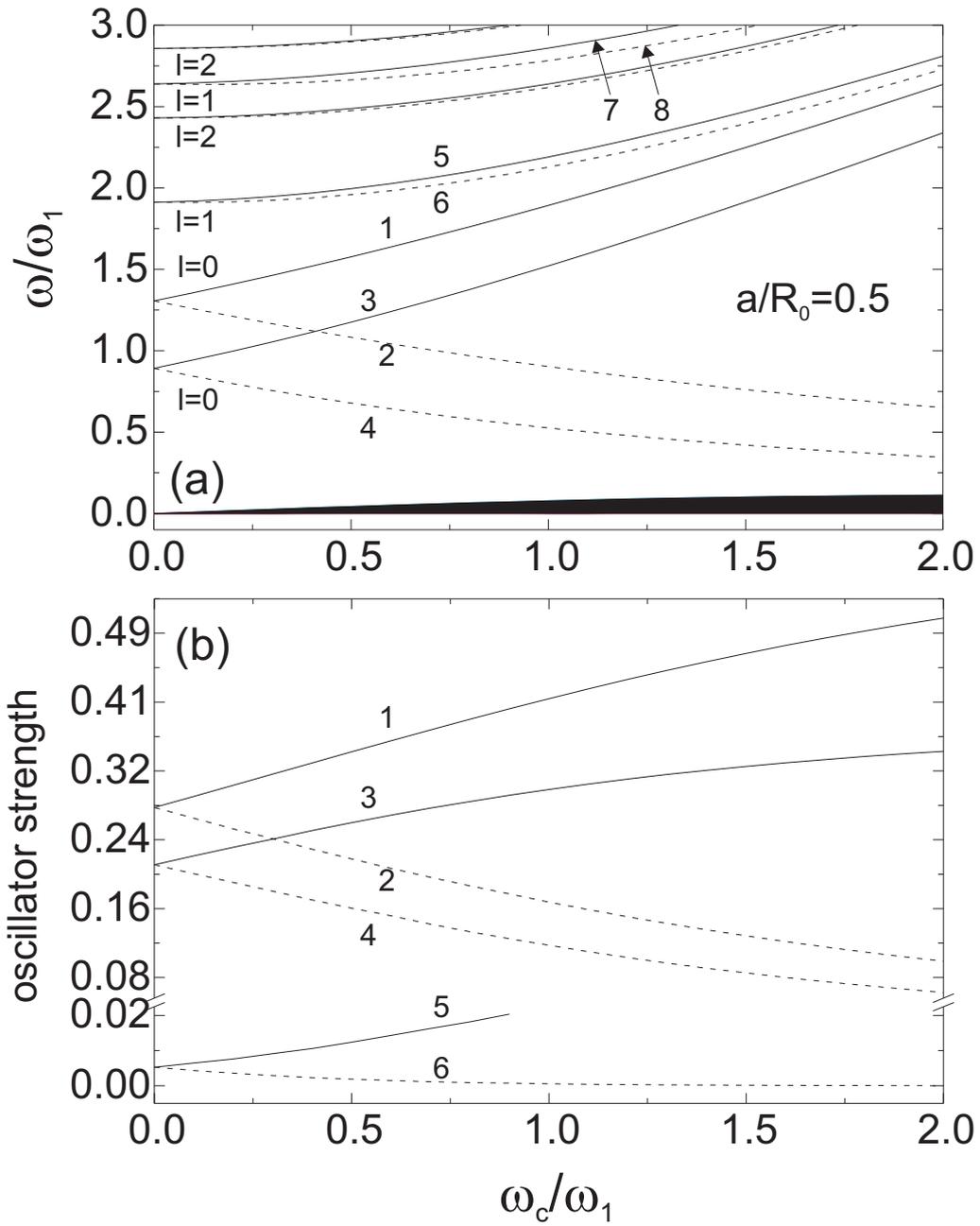
\begin{figure}
\caption{The frequency spectrum and the oscillator strengths for the system with
$\kappa=2$ and $\mu_1=\mu_2$  for the interdot distance $a/R_0=0.5$. The numbers
give the corresponding frequency branches and oscillator strengths. Only 6
oscillator strengths are shown.}
\label{figfreqk2d05}
\end{figure}
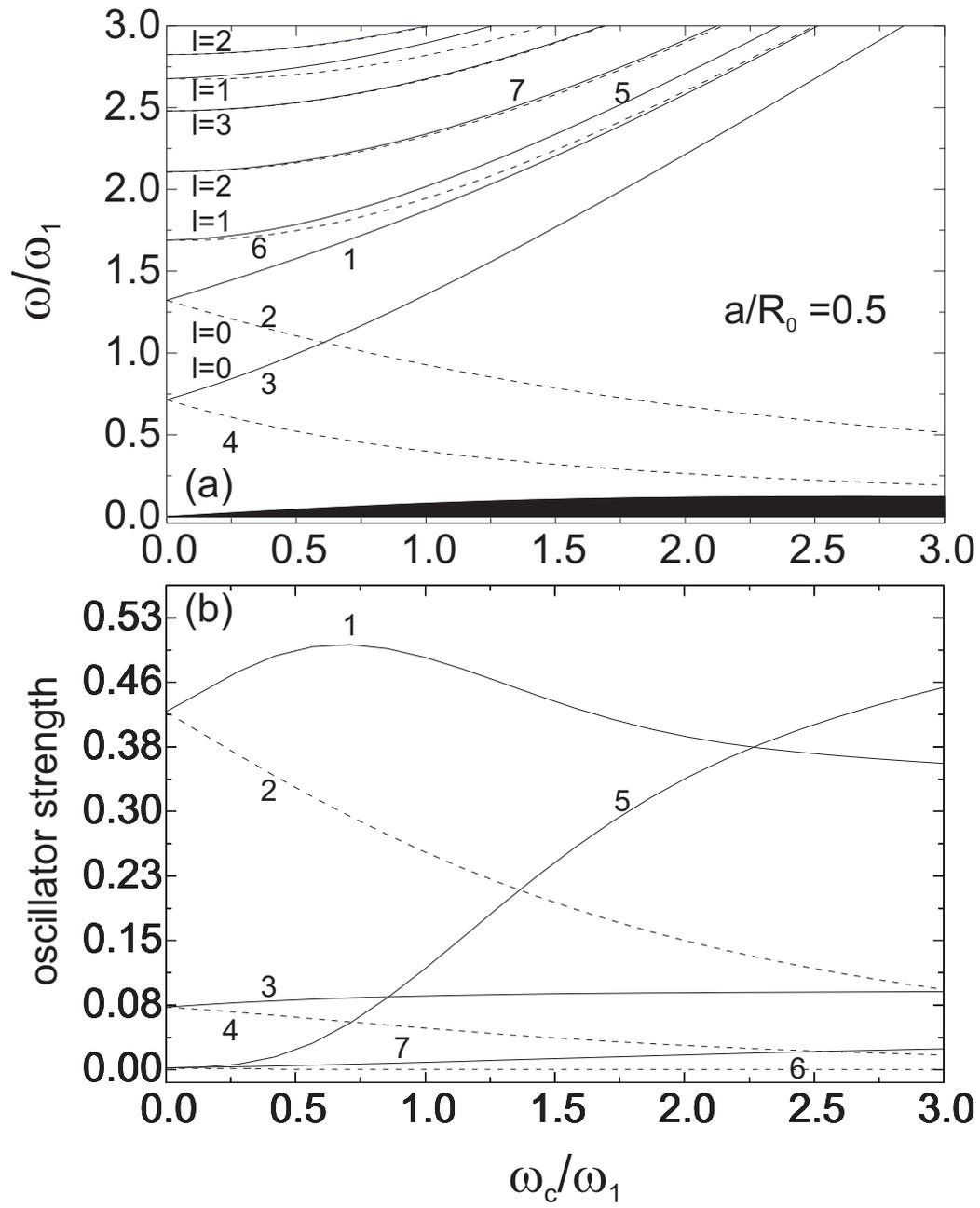
\begin{figure}
\caption{The frequency spectrum and the oscillator strength for the system with
$\kappa=2$, $a/R_0=0.3$ and with an external applied bias 
potential $\mu_1-\mu_2=-0.2$. The numbers
give the corresponding frequency branches and oscillator strengths. Only 7
oscillator strengths are shown.}
\label{figk2mu2}
\end{figure}
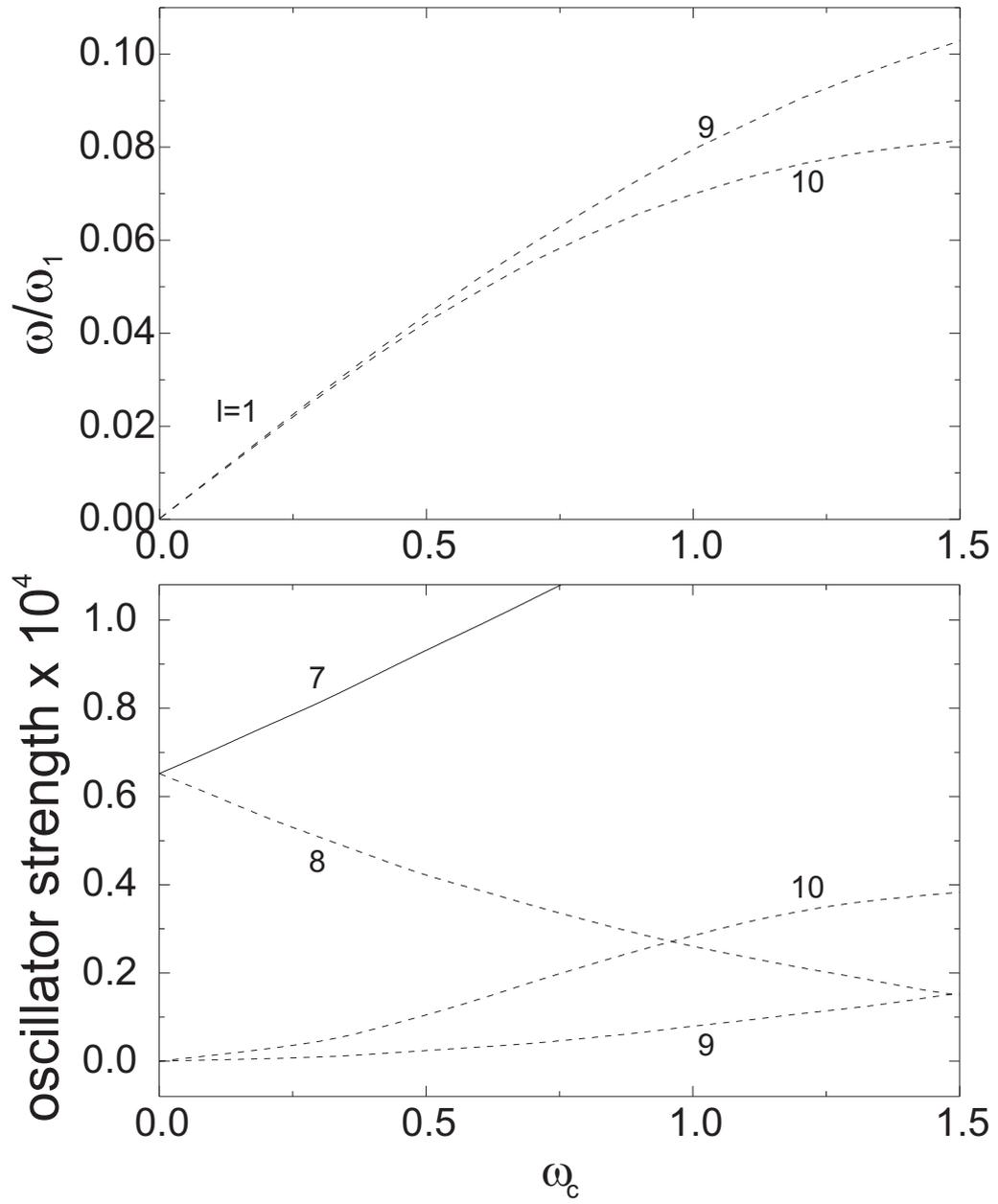
\begin{figure}
\caption{The edge modes and the oscillator strengths for the system with
$\kappa=2$ and $\mu_1=\mu_2$. The interdot distance is $a/R_0=0.5$. The
series~(\ref{pot_dot}) and~(\ref{hpot_dot}) are approximated by only the 
first two terms. The numbers
give the corresponding frequency branches and oscillator strengths.}
\label{figedge}
\end{figure}

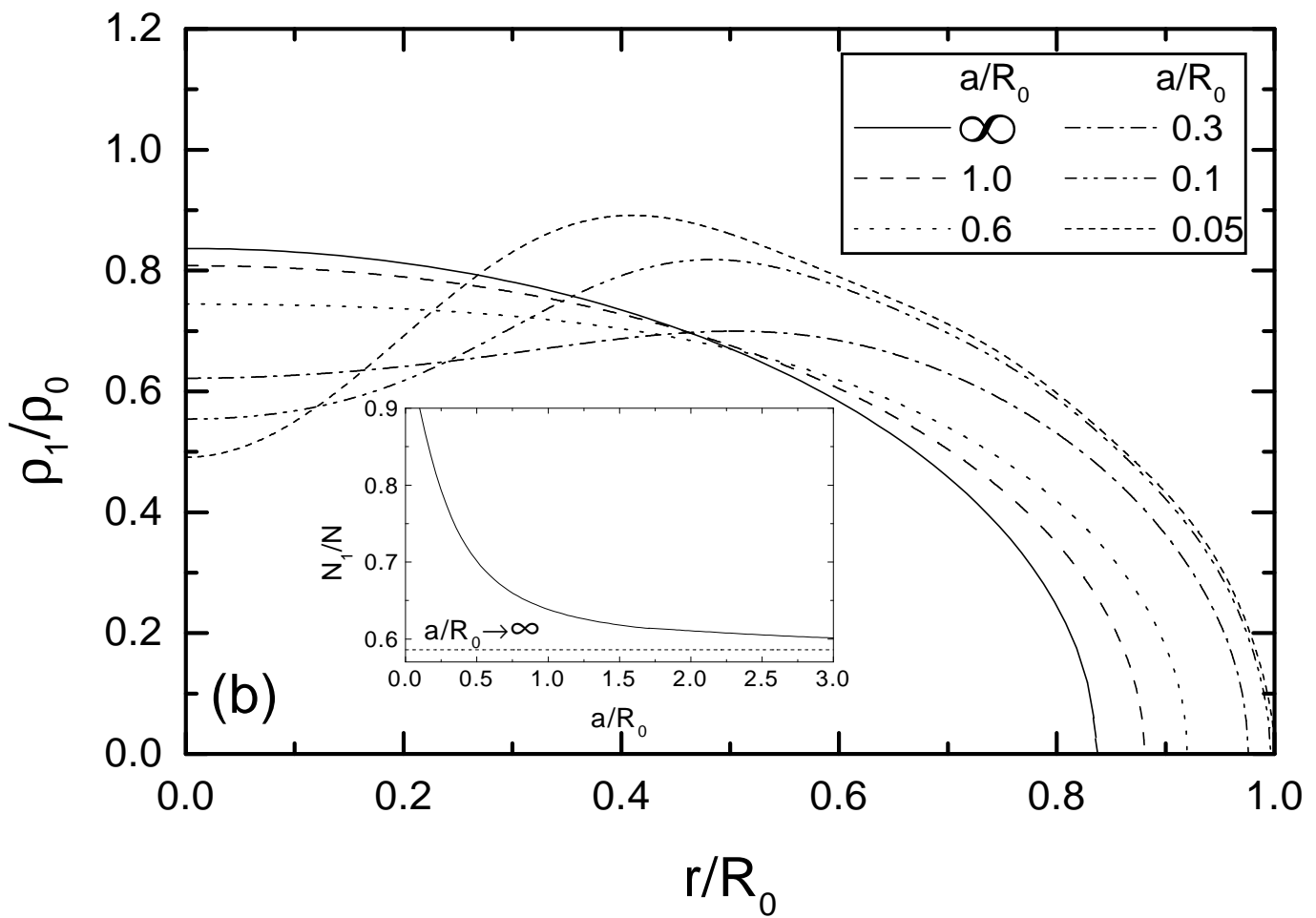
\end{document}